**Volodymyr L. Buriachok**
DSc in Technical Sciences, Professor, Head of the Department of Information and Cyber Security
Borys Grinchenko Kyiv University, Kyiv, Ukraine
ORCID ID: 0000-0002-4055-1494
*v.buriachok@kubg.edu.ua*

**Volodymyr Yu. Sokolov**
PhD in Technical Sciences, associate professor of the Department of Information and Cyber Security
Borys Grinchenko Kyiv University, Kyiv, Ukraine
ORCID ID: 0000-0002-9349-7946
*v.sokolov@kubg.edu.ua*

**Mahyar Taj Dini**
MSc, senior lecturer of the Department of Information and Cyber Security
Borys Grinchenko Kyiv University, Kyiv, Ukraine
ORCID ID: 0000-0001-8875-3362
*m.tajdini@kubg.edu.ua*

# RESEARCH OF CALLER ID SPOOFING LAUNCH, DETECTION, AND DEFENSE

**Abstract.** Caller ID parodying produces the valid Caller character, in this manner deciding seem to start from another client. This apparently basic assault strategy has been utilized in the developing communication fake and trick calls, bringing about significant financial trouble. Unfortunately, callerID spoofing is easy to implement but yet it is difficult to have protection against it. In addition, there are not effective and defense solutions available right now. In this research it is suggested the CIVE (Callee Inference & VErification), a compelling and viable guard against Caller ID spoofing. This way it is described how it's possible to lunch call spoofing and between line describe how CIVE approach method can help to prevent somehow this kind of attacks. Caller ID Spoofing could cause huge financial and political issues special nowadays, when many things even sometimes authentication and verification are available by phone call, like banks approving transactions or two factor authentications and many other things. We believe critical industries specially banks and payment service providers should be protected against such vulnerabilities with their system and make an approach to prevent it, also it is very important to learn people specially who has special social place like politicians or celebrities to know such kind of attack are already exist. For this paper we implemented a call from white house to show there is no limitation and no matter whom you try to spoof, but destination which is the victim receive the call and that make this attack vector dangerous. And even modern communication and even devices like 4G and smart phones are not able to prevent or even detect this kind of attack. This study is a demonstration of the vulnerabilities available. All experiments were conducted on isolated mock-ups.

## 1. INTRODUCTION

Caller ID spoofing is uncomplicated to lunch nevertheless difficult to detect and defend from, even in 4G LTE networks. This work is connected with previous studies on the work with wireless systems presented in [1]–[7].

Fig. 1 portrays a conventional call setup stream for any call innovation. Call signaling runs first to build up a call session and afterward begins voice discussions once again the session. The signaling begins with a setup ask from the Caller to the callee, trailed by all signaling required by the call setup system. The two gatherings acquire call service from their own career network (CN). CNs are co-associated so that call parties from various CNs can converse with each other. Considering the callee's CN as the 4G LTE system. 4G supports two voice arrangements: Voice-over-LTE (VoLTE) [8] and Circuit Switched Fall Back (CSFB) [9]. VoLTE adopts Voice-over-IP (VoIP) and conveys voice calls (and its signaling) in IP bundles; CSFB use inheritance 3G/2G systems to initiate a CS voice calls. Both theoretically bolster comparative signaling however utilize diverse conventions. VoLTE utilizes Session Initiation Protocol (SIP) [10], while CSFB utilizes Call Control (CC) [11]. In spite of the fact that they talk distinctive convention languages, e. g., the primary demand via INVITE in SIP for VoLTE and via SETUP in CC for CSFB/CS, the interpretation is taking care of their border gateway for co-working. For example, INVITE is mapped into SETUP once leaving 4G and entering 3G/2G. Each call party has an all-inclusive exceptional ID, frequently a phone number (e. g., +1 xxx-xxx-xxxx). ID engages as a permanent address-of-record which is doled out upon membership and is verified before use. In particular, cell systems run Authentication and Key Agreement (AKA), which uses the shared secret key, which is hold by SIM (locally) and known just by the administrator (client database) to validate one another.

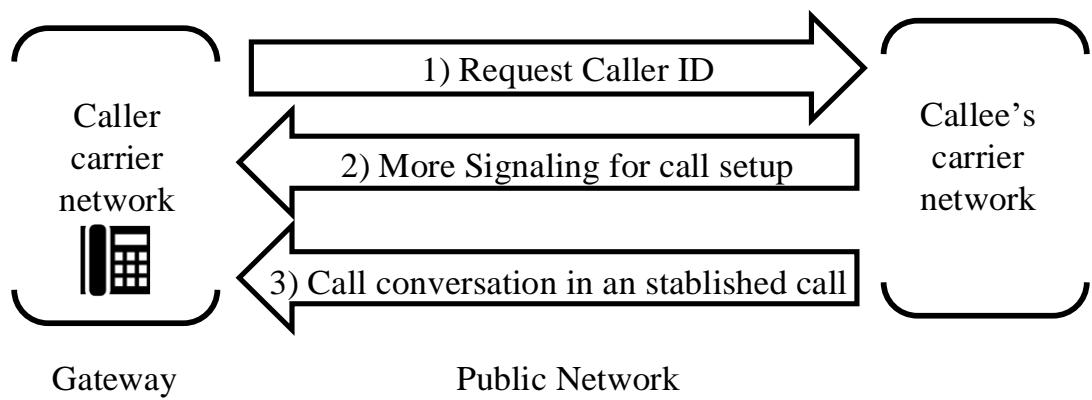

*Fig. 1. Conventional call setup stream*

## 2. CALLER ID TESTS

Caller ID spoofing uses a counterfeit ID. In this research, it was considered the parodying scenario where Caller Ellen (E, in the future) calls the victim Brian (B) by fabricating Alex's ID (A.ID). In reality, Caller ID mocking is in fact plausible and simple. Spoofer E only adjusts the Caller ID conveyed in the configuration request, which is permitted where E's CN does not implement the forwarded Caller ID if it is an equivalent to the authenticated one. Use VoIP/VoLTE for instance.

By using the header 'From' in the INVITE message to count the Caller ID. E places A.ID instead of E.ID, with the goal that B just observes an approaching call from 'A.' The worst thing which is happening is that the caller ID spoofing offered as one open administration by phony ID suppliers To utilize it, E just needs to include B's phone number as the objective one and A's telephone number as the ideal fake Caller ID. Spoofing is free to use and takes a couple of seconds to run it after implementation.

Fig. 2 demonstrates how clients can in simple way use spoofing like an attacker. Local SIP server which truncated with another SIP service provider over the internet and effectively make spoofing assaults towards our test phones (HUAWEI P9) is used. A similar ID was utilized in late genuine trick call, which has brought about more than $1,000,000 loss for a targeted individual [12–15]. Likewise, it is seen that the answer for fighting Caller name spoofing, e. g., True-Caller [16] and Google's dialer [17], probably won't work. Both neglected to distinguish when the genuine trick call occurred in Jan. 2018 (see Fig. 2), however it was published for a half year back [18, 19]. In our research of fake tests 20 days after the fact, TrueCaller attempted to certain degree for the telephone with Internet intern access, but fizzled at the telephone without Internet intern access. In our controlled trials (the both aggressors and victims hold by both researchers), are tested by creating other telephone numbers (mobile or landline, individual or business, from various states and countries, >100 altogether). The research made confirms that, all are easy to spoof with no indication of limitations.

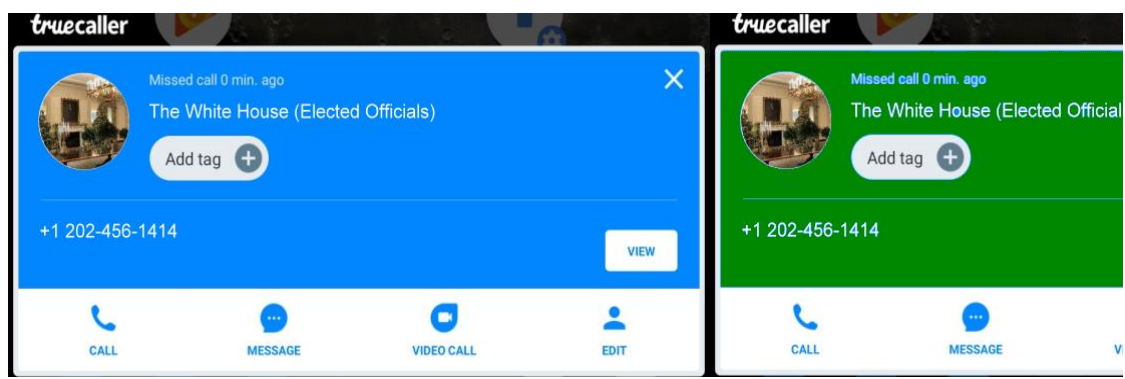

*Fig. 2. Screenshots of the callee phones (Huawei p9) in a real scam and two controlled caller ID spoofing tests*

### 3. BASIC IDEA AND FEASIBILITY OF THE STUDY

The essential thought in Fig. 3 confirms whether the caller ID (A.ID) is a spoofer or not, by looking at the call states of two call sessions. For an approaching call inCall, CIVE asks the callee (i. e., B) to make an auCall back to the starting ID (1). B uses the inCall's setting to surmise the condition of Caller X (An or E in the nonappearance/nearness of spoofing). For instance, X is dialing when in Call rings. In the meantime, B utilizes its own perception on auCall to derive A's call state (2), and evaluate it with X's (3). In case of an inconformity, and is attested to be not X, and spoofing to inCall. The above basic arrangement idea has a few decent highlights. No control is expected on different parts (the carrier infrastructure or different gadgets). It doesn't require participation by others or additional data intern access.

It also works under two terms: (1) B's perception can gather A's particular call state. In the moment when the call state of auCall. Callee changes, the state of auCall. Caller ought to be changed too for making the derivation possible (2).

The interfered A's call state has to contrast from the genuine call state at any rate once upon spoofing. Further possible tests to address a key technical issue shows: which accessible data from the auCall. Caller side can be used to construe the particular state on the remote auCall. Firstly, basic call data provided by mobile OSes (utilizing Android as an instance) something like PRECISE_CALL_STATE, PHONE_STATE in telephony manager and framework logs were exanimated. Nonetheless, it is presumed that such data neglects to deduce the state on the remote callee side, because it just gives call states alone side. Particularly, the two sides are in a similar call session, and the Caller ought to almost certainly realize what occurs at the ending part. By the way, in practice these abnormal state APIs cover up fine-grained call setting and fail the run surmising, required by CIVE. In such way there have been looked for pure context call setting data. While experiments it was set out that, the succession of call setup signaling messages (SIP for VoLTE and CC for CSFB/CS) comply with expectations.

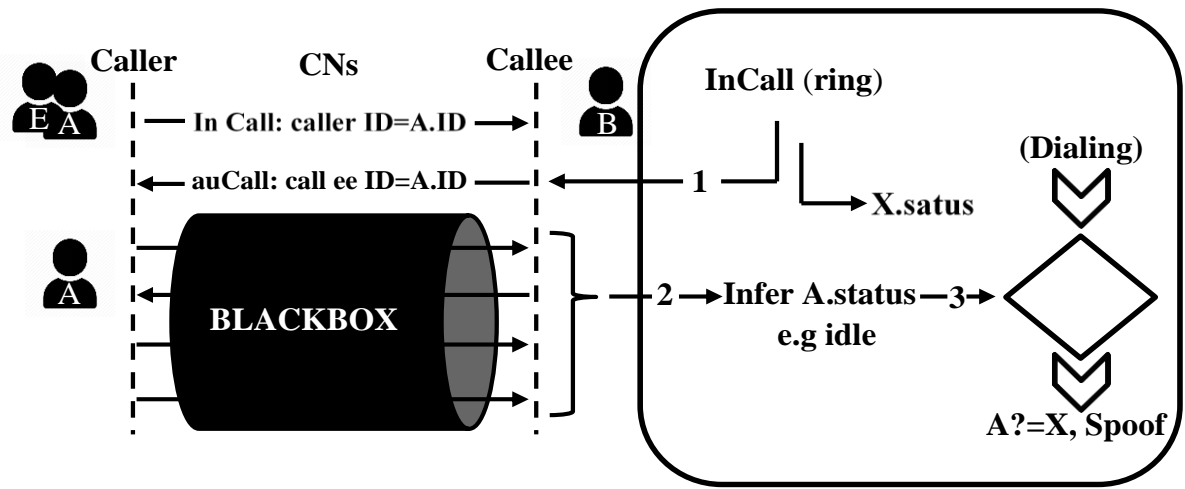

*Fig. 3. Sequencing detection whether the caller ID (A.ID) is a spoofer or not*

## 4. BASELINE FEASIBILITY TESTS

Initially essential possibility tests to approve that the call signaling messages received by the Caller's side are sufficient to deduce the callee's call state have been run, moreover run out the trials in three ordinal call settings:

- (C1) A calls B (no-spoof)
- (C2) E calls B while A is idle (spoof-idle)
- (C3) E calls B while A is on a call (spoof-conn)

SIP signaling messages for auCall utilizing **tcp-dump** or wireshark at telephone B, an established Android gadget were gathered. 10+ telephone models (from Samsung, Google, Huawei, Sony, Xiaomi, and so on) and found no distinction have been attempted with.

They were also tested with all of the four best level US carriers: AT&T, T-Mobile, Verizon and Sprint. The studied material differences incidentally, while all are demonstrated feasible (Fig. 4). The T-Mobile results to delineate CIVE's possibility were utilized.

| | Time ▲ | Source | Destination | Protocol | Length | Info |
|---|---|---|---|---|---|---|
| 2 | 0.00… | 2607:fc… | fd00:97… | SIP/SDP | 984 | Request: INVITE sip:+1▇ |
| 5 | 0.38… | fd00:97… | 2607:fc… | SIP | 496 | Status: 100 Trying \| |
| 9 | 2.79… | fd00:97… | 2607:fc… | SIP/SDP | 332 | Status: 183 Session Prog… |
| 11 | 2.87… | 2607:fc… | fd00:97… | SIP | 884 | Request: PRACK sip:sgc_c( |
| … | 4.22… | fd00:97… | 2607:fc… | SIP | 984 | Status: 180 Ringing \| |

```
▼ Status-Line: SIP/2.0 180 Ringing
   ▶ To: <sip:▇▇▇@s▇▇c.t-mobile.com;use
   ▶ From: <sip:1▇▇▇@m▇▇c.t-mobile.com>;1
   ▶ Record-Route: <sip:[FD00:976A:C206:1821::1]:6
      P-Charging-Vector: icid-value=sgc11.nvatf002.
      P-Early-Media: sendrecv
   ▶ Feature-Caps: *;+g.3gpp.srvcc
      Alert-Info: <urn:alert:service:call-waiting>
```

| | | | | | | |
|---|---|---|---|---|---|---|
| 23 | 10.8… | 2607:fc… | fd00:97… | SIP | 864 | Request: CANCEL sip:+1▇ |
| 25 | 10.9… | fd00:97… | 2607:fc… | SIP | 572 | Status: 200 OK \| |
| 27 | 11.0… | fd00:97… | 2607:fc… | SIP | 788 | Status: 487 Request Term… |
| 33 | 11.1… | 2607:fc… | fd00:97… | SIP | 684 | Request: ACK sip:+1▇▇▇9 |

*Fig. 4. Multi variable baseline feasibility test findings*

Fig. 5 plots the charts of SIP signaling messages observed at B in C1-C3 scenarios. auCall is started, when inCall rings yet isn't recognized by B. Three objective facts were separated.

The First one: the sequences of call signaling messages share numerous common parts in each of the three situations. Specifically, all begin with INVITE, pursued by100→183→⋯→180⋯→200⋯. These numbers show the SIP state and response codes, which all are standards [20].

The second illustrated that each sequence contains certain basic data to recognize three call settings. For instance, in the received '180 Ringing message,' there are two fields: P-Early-Media (PEM) and Alert Info (detailed signs in Fig. 5). Table 1 lists their particular qualities in each one of the three situations.

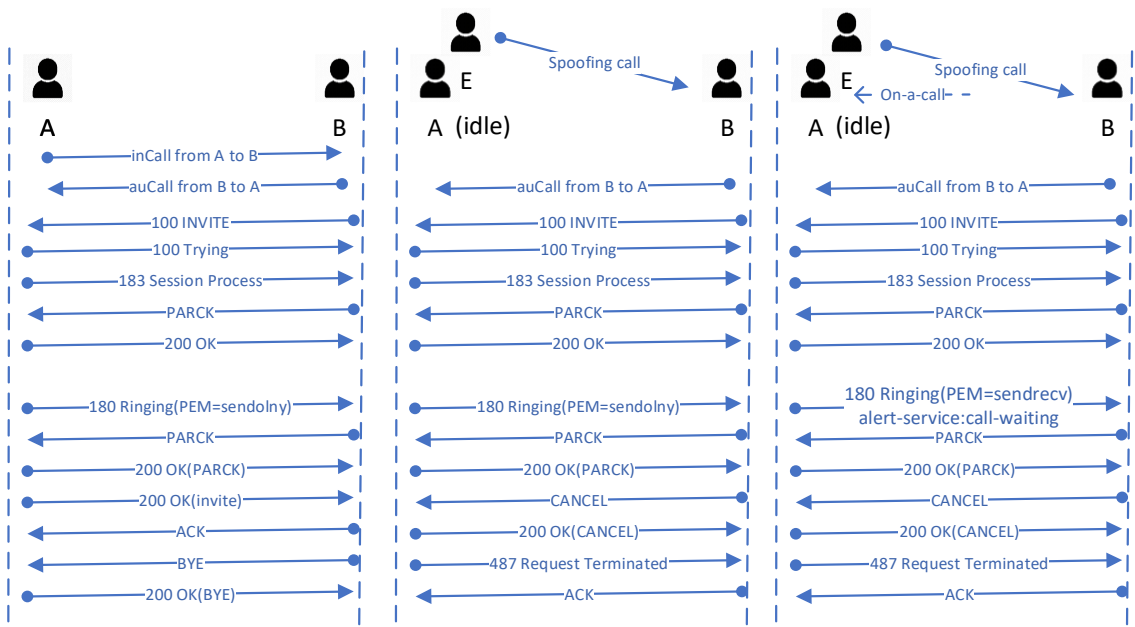

*Fig. 5. Caller ID count research by using the header 'From' in the INVITE message*

The third, additionally was found repetitive features which can induce distinct call state also C1 observes 200 but C2/C3 uses '487 Request Terminated,' responding to INVITE; C1 utilizes BYE while C2/C3 uses CANCEL at the end. Therefore, it is abused the unexplored side channel of call setup signaling messages. As a result, particular callee state: dialing (C1), idle (C2), and conn (C3) were gathered, while the final state in C2/C3 contrasts from the anticipated state without spoofing (C1).

*Table 1*

**Caller trials in three ordinal call settings**

|      | True state   | Key observations (features)              |
|------|--------------|------------------------------------------|
| C1   | A is dialing | 180.PEM= sendonly                        |
| C2   | A is idle    | 180.PEM= sendrecv                        |
| C3   | A is conn    | 180.PEM= sendrecv, 180.ALERT= call-waiting |

## 5. CLARIFICATION OF ARRANGEMENT EFFICIENCY

Now it will clarified why the above arrangement should work. The main reason lies in the call setup technique modeled by Internet RFCs and cellular determinations. Table 2 exemplifies such vital information. First, call setup signaling messages contain detailed or latent data, which is related to the callee's state to facilitate the call setup.

The call request can come to the callee if (s)he is available. In case the callee is occupied, the research can be dropped from the ringtone or being changed to the voice mailbox. If the callee has call pausing, the request call in any case can contact him/her if (s)he is in a call.

Second, standard terms command a rich arrangement of signaling messages, which convey rich setting data and can be abused to induce the call state on the other party. SIP defines numerous parameters and response codes. For example, '180 Ringing' indicates that the call request reached the callee; '181 Call Is Being Forwarded' is utilized when the call is sent to a voice mailbox for an occupied callee; '486 Busy Here' demonstrates an occupied callee. In addition, SIP characterizes extensions to pass more information.

For instance, the P-Early-Media (PEM) field authorizes like an early media (e. g., ringtone), with 'sendrecv' indicating bidirectional line, 'sendonly,' 'recvonly,' and 'inactive' indicating a directional line to the Caller, from the Caller, and no line. Another example is URN-Alert (Fig. 5), which gives ordinary understandings of the referenced tones. 'Call-pausing' demonstrates that the callee is in an active or held call, and 'forward' shows the call will be forwarded. Third, these signaling messages are related with the call setup's limited state machine (FSM), which together shows more call state data.

*Table 2*

**Exemplification of call setup technique modeled by Internet RFCs and cellular determinations**

| Field | References and values |
|---|---|
| SIP response codes | RFC3261 [21]:<br>• 200—ok<br>• 180—ringing<br>• 181—call is being forwarded<br>• 182—queued<br>• 183—session progress<br>• 301—moved permanently<br>• 480—temporarily unavailable<br>• 481—call/transaction does not exist<br>• 486—busy here<br>• 487—request terminated |
| PEM | RFC5009 [19]:<br>• sendrecv<br>• sendonly<br>• recvonly<br>• inactive |
| URN-Alert | RFC7462 [20]:<br>• normal (default)<br>• call-waiting<br>• forward<br>• recall:callback<br>• recall:hold<br>• recall:transfer |
| VoLTEFSM | TS24.229 [22], TS24.628 [23], and TS24.615 [24]:<br>carrying early-media value or alert-info in 180/183,<br>call terminated by network when busy |

As an example, the request '487 Request Terminated' suggests the demand to be finished by a BYE or CANCEL request [23]. The Caller sends CANCEL when intending to finish a call before the call is replied. It sends BYE if the first INVITE still returns '200 OK.' CANCEL is seen without '200 OK' as a response of INVITE. Cellular terms [22–24] additionally declare that VoLTE receives certain signaling messages, which are helpful for callee state interference from the Caller side observations. In outline, call setup utilizes a Stateful FSM and its signaling likely collects enough data to conclude callee state. This makes callee state inference dependent on the signaling message arrangement, saw on the Caller side.

## 6. CONCLUSION AND FUTURE WORK

All industries specially banking system should check this vulnerability into their system and make an approach to prevent it, also it is very important to learn people specially who has special social place like politicians or celebrities to know such kind of attack are already exist and don't be scam by same method.

The research exhibits the plan and assessment of CIVE while real Caller ID spoofing attack lunched and described how easy possible to change call header information. Therefore, it introduces a non-typical, just callee arrangement against Caller ID spoofing. It devises different derivation procedures to construe the remote Caller state, by analyzing an unexplored side channel of 4G networks.

We are looking for a better solution to protect against Caller-ID spoofing with detection method, on victim (call destination) side, because we believe even if there be mitigation methods by providers, still possible to be a malicious provider, but in such cases best solution to move main layer of protection to end-point.

## REFERENCES

[1] V. M. Astapenya and V. Y. Sokolov, "Modified accelerating lens as a means of increasing the throughput, range and noise immunity of IEEE 802.11 systems," in *2015 International Conference on Antenna Theory and Techniques (ICATT)*, Apr. 2015. https://doi.org/10.1109/ICATT.2015.7136852.

[2] V. M. Astapenya and V. Yu. Sokolov, "Experimental evaluation of the shading effect of accelerating lens in azimuth plane," in *2017 XI International Conference on Antenna Theory and Techniques (ICATT)*, pp. 388–390, 2017. https://doi.org/10.1109/ICATT.2017.7972671.

[3] V. Sokolov, A. Carlsson, I. Kuzminykh, "Scheme for dynamic channel allocation with interference reduction in wireless sensor network," in *2017 4th International Scientific-Practical Conference Problems of Infocommunications. Science and Technology (PIC S&T)*, pp. 564–568, 2017. https://doi.org/10.1109/INFOCOMMST.2017.8246463.

[4] I. Bogachuk, V. Sokolov, and V. Buriachok, "Monitoring Subsystem for Wireless Systems based on Miniature Spectrum Analyzers," in *2018 International Scientific-Practical Conference Problems of Infocommunications. Science and Technology (PIC S&T)*, Oct. 2018. https://doi.org/10.1109/infocommst.2018.8632151.

[5] V. Y. Sokolov, "Comparison of Possible Approaches for the Development of Low-Budget Spectrum Analyzers for Sensory Networks in the Range of 2.4–2.5 GHz," *Cybersecurity: Education, Science, Technique*, no. 2, pp. 31–46, 2018. https://doi.org/10.28925/2663-4023.2018.2.3146.

[6] M. Vladymyrenko, V. Sokolov, and V. Astapenya, "Research of Stability in Ad Hoc Self-Organized Wireless Networks," *Cybersecurity: Education, Science, Technique*, no. 3, pp. 6–26, 2019. https://doi.org/10.28925/2663-4023.2019.3.626.

[7] V. Sokolov, B. Vovkotrub, and Y. Zotkin, "Comparative Bandwidth Analysis of Lowpower Wireless IoT-Switches," *Cybersecurity: Education, Science, Technique*, no. 5, pp. 16–30, 2019. https://doi.org/10.28925/2663-4023.2019.5.1630.

[8] GSM Association. (2015). "Voice over LTE." [Online]. Available: http://www.gsma.com/technicalprojects/volte/ [Sep. 30, 2019].

[9] *Circuit Switched (CS) Fallback in Evolved Packet System (EPS)*, TS23.272, 2017.

[10] J. Rosenberg, et. al. (2002). "RFC3261: SIP: Session Initiation Protocol." [Online]. Available: https://tools.ietf.org/html/rfc3261 [Sep. 30, 2019].

[11] *Mobile Radio Interface Signalling Layer 3. General Aspects*, TS24.007, 2011.

[12] M. Xuequan. (2017). "Chinese Police Arrest 118 in Scam Targeting Seniors." [Online]. Available: http://www.xinhuanet.com/english/2017-09/20/c_136624766.htm [Sep. 28, 2019].

[13] Phoenix New Media. (2018). "Alert! Phone Scam Targeting Chinese from China's Consulates across the US! Someone Lost Millions of Dollars" (in Chinese). [Online]. Available: http://wemedia.ifeng.com/47830827/wemedia.shtml [Sep. 29, 2019].

[14] Consulate General of the People's Republic of China in New York. (2017, Apr.). "Phone Scam Alert." [Online]. Available: http://newyork.china-consulate.org/eng/lqfw/lsbhyxz/t1486921.htm [Sep. 30, 2019].

[15] Xinhua. (2017). "Phone Scams Targeting NYC Chinese Communities Exposed." [Online]. Available: http://www.xinhuanet.com/english/2017-08/10/c_136513524.htm [Sep. 28, 2019].

[16] True Software Scandinavia. (2017). "TrueCaller." [Online]. Available: https://www.truecaller.com [Sep. 27, 2019].

[17] Google. (2016). "Google Phone App." [Online]. Available: https://play.google.com/store/apps/details?id=com.google.android.dialer [Sep. 30, 2019].

[18] Android. (2017). "Telephony Manager." [Online]. Available: https://developer.android.com/reference/android/telephony/TelephonyManager.html [Sep. 30, 2019].

[19] *Private Header (P-Header) Extension to the Session Initiation Protocol (SIP) for Authorization of Early Media*, RFC5009, 2007.

[20] *URNs for the Alert-Info Header Field of the Session Initiation Protocol (SIP)*, RFC7462, 2015.

[21] *SIP: Session Initiation Protocol*, RFC3261, 2002.

[22] *IP Multimedia Call Control Protocol based on Session Initiation Protocol (SIP) and Session Description Protocol (SDP). Stage 3*, TS24.229, 2017.

[23] *Common Basic Communication Procedures using IP Multimedia (IM) Core Network (CN) Subsystem*, TS24.628, 2017.

[24] *Communication Waiting (CW) using IP Multimedia (IM) Core Network (CN) subsystem; Protocol Specification*, TS24.615, 2017.

**Бурячок Володимир Леонідович**
д.т.н., професор, завідуючий кафедри інформаційної та кібернетичної безпеки
Київський університет імені Бориса Грінченка, Київ, Україна
ORCID ID: 0000-0002-4055-1494
*v.buriachok@kubg.edu.ua*

**Соколов Володимир Юрійович**
к.т.н., доцент кафедри інформаційної та кібернетичної безпеки
Київський університет імені Бориса Грінченка, Київ, Україна
ORCID ID: 0000-0002-9349-7946
*v.sokolov@kubg.edu.ua*

**Таджіні Махіяр**
старший викладач кафедри інформаційної та кібернетичної безпеки
Київський університет імені Бориса Грінченка, Київ, Україна
ORCID ID: 0000-0001-8875-3362
*m.tajdini@kubg.edu.ua*

# ДОСЛІДЖЕННЯ СПУФІНГУ ІДЕНТИФІКАТОРА АБОНЕНТА ПРИ РЕЄСТРАЦІЇ: ВИЯВЛЕННЯ ТА ПРОТИДІЯ

**Анотація.** При підробленні ідентифікатора абонента надається дійсний доступ до сервісів від імені іншого абонента. Ця основна стратегія нападу часто застосовується в існуючих телекомунікаційних мережах для підробки и фальсифікації доступу, що спричиняє значні фінансові збитки. Нажаль, підробку ідентифікатора абонента здійснити досить легко, а захиститися від неї вкрай важко. Крім того, зараз не існує ефективних рішень по протидії цій вразливості. У цьому дослідженні пропонується застосування CIVE (Callee Inference & VErification) — можливий захист від підроблення ідентифікатора абонента. В статті представлено, як можна підробляти виклики, а також представлений метод CIVE, за допомогою якого можливо частково протидіяти подібним нападам. Підроблення ідентифікатора абонента може спричинити величезні фінансові та політичні проблеми, особливо сьогодні, коли багато речей, навіть автентифікація та підтвердження, доступні за допомогою телефонного дзвінка, наприклад, при доступі до банківських рахунків, підтвердження транзакцій за допомогою двофакторної автентифікації та багато інших речей. Ми вважаємо, що у критичних галузях, зокрема для банків та постачальників інших платіжних послуг, інформаційні системи повинні бути захищені від таких вразливостей і мати на озброєні методи запобігання. Також важливо навчити персонал, який має особливе соціальне значення — політиків і знаменитостей, що такі види атака вже існують і можуть призвести до іміджевих втрат. Для цього документу ми реалізували дзвінок з Білого дому, щоб показати відсутність обмежень при виборі жертви, а потерпілий отримує дзвінок і робить цей вектор нападу небезпечним. Ані сучасні телекомунікаційні компанії, ані виробники пристроїв на зразок 4G смартфонів не здатні запобігти або навіть виявити подібний напад. Це дослідження є демонстрацією наявних уразливих місць. Усі експерименти проводилися на ізольованих макетах.

**Keywords:** ідентифікатор абонента; спуфінг; CIVE; визначення абонента; перевірка абонента; абонент; SIP; протокол ініціювання сесії.

## REFERENCES

[1] V. M. Astapenya and V. Y. Sokolov, "Modified accelerating lens as a means of increasing the throughput, range and noise immunity of IEEE 802.11 systems," in *2015 International Conference on Antenna Theory and Techniques (ICATT)*, Apr. 2015. https://doi.org/10.1109/ICATT.2015.7136852.

[2] V. M. Astapenya and V. Yu. Sokolov, "Experimental evaluation of the shading effect of accelerating lens in azimuth plane," in *2017 XI International Conference on Antenna Theory and Techniques (ICATT)*, pp. 388–390, 2017. https://doi.org/10.1109/ICATT.2017.7972671.

[3] V. Sokolov, A. Carlsson, I. Kuzminykh, "Scheme for dynamic channel allocation with interference reduction in wireless sensor network," in *2017 4th International Scientific-Practical Conference Problems of Infocommunications. Science and Technology (PIC S&T)*, pp. 564–568, 2017. https://doi.org/10.1109/INFOCOMMST.2017.8246463.

[4] I. Bogachuk, V. Sokolov, and V. Buriachok, "Monitoring Subsystem for Wireless Systems based on Miniature Spectrum Analyzers," in *2018 International Scientific-Practical Conference Problems of Infocommunications. Science and Technology (PIC S&T)*, Oct. 2018. https://doi.org/10.1109/infocommst.2018.8632151.

[5] V. Y. Sokolov, "Comparison of Possible Approaches for the Development of Low-Budget Spectrum Analyzers for Sensory Networks in the Range of 2.4–2.5 GHz," *Cybersecurity: Education, Science, Technique*, no. 2, pp. 31–46, 2018. https://doi.org/10.28925/2663-4023.2018.2.3146.

[6] M. Vladymyrenko, V. Sokolov, and V. Astapenya, "Research of Stability in Ad Hoc Self-Organized Wireless Networks," *Cybersecurity: Education, Science, Technique*, no. 3, pp. 6–26, 2019. https://doi.org/10.28925/2663-4023.2019.3.626.

[7] V. Sokolov, B. Vovkotrub, and Y. Zotkin, "Comparative Bandwidth Analysis of Lowpower Wireless IoT-Switches," *Cybersecurity: Education, Science, Technique*, no. 5, pp. 16–30, 2019. https://doi.org/10.28925/2663-4023.2019.5.1630.

[8] GSM Association. (2015). "Voice over LTE." [Online]. Available: http://www.gsma.com/technicalprojects/volte/ [Sep. 30, 2019].

[9] *Circuit Switched (CS) Fallback in Evolved Packet System (EPS)*, TS23.272, 2017.

[10] J. Rosenberg, et. al. (2002). "RFC3261: SIP: Session Initiation Protocol." [Online]. Available: https://tools.ietf.org/html/rfc3261 [Sep. 30, 2019].

[11] *Mobile Radio Interface Signalling Layer 3. General Aspects*, TS24.007, 2011.

[12] M. Xuequan. (2017). "Chinese Police Arrest 118 in Scam Targeting Seniors." [Online]. Available: http://www.xinhuanet.com/english/2017-09/20/c_136624766.htm [Sep. 28, 2019].

[13] Phoenix New Media. (2018). "Alert! Phone Scam Targeting Chinese from China's Consulates across the US! Someone Lost Millions of Dollars" (in Chinese). [Online]. Available: http://wemedia.ifeng.com/47830827/wemedia.shtml [Sep. 29, 2019].

[14] Consulate General of the People's Republic of China in New York. (2017, Apr.). "Phone Scam Alert." [Online]. Available: http://newyork.china-consulate.org/eng/lqfw/lsbhyxz/t1486921.htm [Sep. 30, 2019].

[15] Xinhua. (2017). "Phone Scams Targeting NYC Chinese Communities Exposed." [Online]. Available: http://www.xinhuanet.com/english/2017-08/10/c_136513524.htm [Sep. 28, 2019].

[16] True Software Scandinavia. (2017). "TrueCaller." [Online]. Available: https://www.truecaller.com [Sep. 27, 2019].

[17] Google. (2016). "Google Phone App." [Online]. Available: https://play.google.com/store/apps/details?id=com.google.android.dialer [Sep. 30, 2019].

[18] Android. (2017). "Telephony Manager." [Online]. Available: https://developer.android.com/reference/android/telephony/TelephonyManager.html [Sep. 30, 2019].

[19] *Private Header (P-Header) Extension to the Session Initiation Protocol (SIP) for Authorization of Early Media*, RFC5009, 2007.

[20] *URNs for the Alert-Info Header Field of the Session Initiation Protocol (SIP)*, RFC7462, 2015.

[21] *SIP: Session Initiation Protocol*, RFC3261, 2002.

[22] *IP Multimedia Call Control Protocol based on Session Initiation Protocol (SIP) and Session Description Protocol (SDP). Stage 3*, TS24.229, 2017.

[23] *Common Basic Communication Procedures using IP Multimedia (IM) Core Network (CN) Subsystem*, TS24.628, 2017.

[24] *Communication Waiting (CW) using IP Multimedia (IM) Core Network (CN) subsystem; Protocol Specification*, TS24.615, 2017.